\documentclass[pra,aps,twocolumn,showpacs]{revtex4}

\usepackage{amsmath}
\usepackage{bm}
\usepackage{graphicx}

\begin{document}

\title{Topological defect formation in quenched ferromagnetic Bose-Einstein
condensates}

\author{Hiroki Saito$^1$}
\author{Yuki Kawaguchi$^2$}
\author{Masahito Ueda$^{2,3}$}
\affiliation{$^1$Department of Applied Physics and Chemistry, The
University of Electro-Communications, Tokyo 182-8585, Japan \\
$^2$Department of Physics, Tokyo Institute of Technology,
Tokyo 152-8551, Japan \\
$^3$Macroscopic Quantum Control Project, ERATO, JST, Bunkyo-ku, Tokyo
113-8656, Japan
}

\date{\today}

\begin{abstract}
We study the dynamics of the quantum phase transition of a ferromagnetic
spin-1 Bose-Einstein condensate from the polar phase to the
broken-axisymmetry phase by changing magnetic field, and find the
spontaneous formation of spinor domain walls followed by the creation of
polar-core spin vortices.
We also find that the spin textures depend very sensitively on the initial
noise distribution, and that an anisotropic and colored initial noise is
needed to reproduce the Berkeley experiment [Sadler {\it et al.}, Nature
{\bf 443}, 312 (2006)].
The dynamics of vortex nucleation and the number of created vortices
depend also on the manner in which the magnetic field is changed.
We point out an analogy between the formation of spin vortices from domain
walls in a spinor BEC and that of vortex-antivortex pairs from dark
solitons in a scalar BEC.
\end{abstract}


\maketitle

\section{Introduction}

Topological defects have played a key role in understanding the physics of
scalar and spinor superfluids.
The first topological defect observed in a gaseous Bose-Einstein
condensate (BEC) was a quantized vortex created using the phase imprinting
method~\cite{Matthews}.
Vortices have also been created using rotating potential~\cite{Madison}
and by means of adiabatic spin rotation accompanied by a topological Berry
phase~\cite{Leanhardt}.
In these methods, topological defect formation is enforced by an external
laser or magnetic field.
On the other hand, topological defects can also be created spontaneously.
The Mermin-Ho texture~\cite{Mermin} of superfluid $^3{\rm He}$ in a
cylindrical container involves a coreless vortex.
The Kibble-Zurek mechanism~\cite{Kibble,Zurek} in a quenched superfluid or
in the early universe affords another intriguing example of spontaneous
topological defect formation.
Thermally nucleated vortices have recently been observed in a gaseous BEC,
presenting yet another topological phase transition known as the
Berezinskii-Kosterlitz-Thouless transition~\cite{Hadzibabic}.

\begin{figure}[tb]
\includegraphics[width=8.4cm]{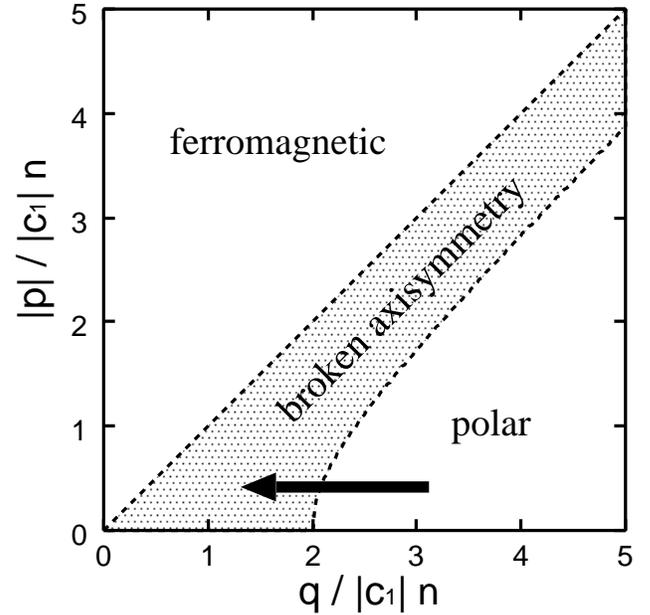}
\caption{
Phase diagram of an homogeneous spin-1 BEC with $c_1 < 0$.
In the ferromagnetic phase, the ground state is the $m = 1$ state for $p >
0$ and the $m = -1$ state for $p < 0$, while in the polar phase, the
ground state is the $m = 0$ state.
The ground-state wave function in the broken-axisymmetry phase possesses
three nonzero components given in Eq.~(\ref{baxi}), and the spin vector
tilts against the applied magnetic field.
The arrow indicates the direction of the quench of the magnetic field.
}
\label{f:phase}
\end{figure}

Recently, the Berkeley group~\cite{Sadler} observed the spontaneous
topological defect formation of spin in a spin-1 $^{87}{\rm Rb}$ BEC.
The atoms were prepared in the $m = 0$ state (polar phase in
Fig.~\ref{f:phase}) in a strong magnetic field, where $m$ is the magnetic
quantum number.
The magnetic field is then lowered below a certain critical value
(broken-axisymmetry phase in Fig.~\ref{f:phase}), where the $m = 0$ state
becomes dynamically unstable and magnetization grows in a direction
perpendicular to the magnetic field, and, consequently, the axisymmetry in
the spin space is spontaneously broken~\cite{Murata}.
In the experiment, the magnetic field was rapidly decreased across the
critical value (as shown by the arrow in Fig.~\ref{f:phase}) --- a process
we refer to as ``quench'' --- and the ensuing magnetization dynamics was
observed by the spin-sensitive {\it in situ} measurement~\cite{Higbie}.
After the quench, it was observed that magnetization grew to form
complicated ferromagnetic domains, and that, remarkably, some snapshots of
the spatial spin distribution revealed topological spin textures, known as
polar-core spin vortices~\cite{Isoshima}.

It has been predicted that a variety of spin textures, such as the
staggered domain and helical textures, are created by the quench of the
magnetic field~\cite{SaitoA1}.
The spontaneous nucleation of the polar-core vortex has also been
predicted in Ref.~\cite{SaitoL}.
The underlying physics in the spontaneous spin-texture formation is spin
conservation, which prohibits uniform magnetization.

Motivated by the Berkeley experiment~\cite{Sadler}, in the present paper
we study the dynamics of a spin-1 BEC caused by the quench of the magnetic
field from the polar to the broken-axisymmetry phase, and investigate the
formation dynamics of the topological spin texture.
We show that the spin vortices observed in Ref.~\cite{Sadler} are formed
in two steps.
First, spin domains separated by domain walls develop, breaking the
axisymmetry in the spin space.
Secondly, the domains transform into spin vortex-antivortex pairs.
We find that the details of the dynamics depend on the initial seed in the
$m = \pm 1$ states, which represents residual atoms, quantum fluctuations,
and thermal noises, and identify an initial seed that reproduces the
experimental results. 
We also study the quench-time dependence of the number of created spin
vortices.

This paper is organized as follows.
Section~\ref{s:meanfield} reviews the mean-field theory of the spin-1
BEC and some of its key properties relevant to later discussions.
We also point out an analogy between defects in a spinor BEC and those in
a scalar BEC in Sec.~\ref{s:topological}.
Section~\ref{s:dynamics} studies the magnetization dynamics of the trapped
system.
Sections~\ref{s:uniform}, \ref{s:noise}, and \ref{s:cutoff} examine three
different kinds of initial noise conditions and identify an appropriate
one that captures the main features of the Berkeley experiment.
Section~\ref{s:quench} investigates the dependence of the dynamics on the
speed of the quench and on the final value of the magnetic field.
Section~\ref{s:transverse} studies the behavior of the growth of the total
magnetization, and Sec.~\ref{s:conclusion} concludes the paper.

\section{Mean-field theory for spin-1 BEC}
\label{s:meanfield}

\subsection{Formulation of the problem}

The zero-temperature mean-field energy of a spinor BEC confined in an
optical trapping potential $V$ is given by
\begin{equation}
E = \int d\bm{r} \left[ \sum_m \psi_m^* \left( -\frac{\hbar^2}{2 M}
\nabla^2 + V \right) \psi_m \right] + E_{\rm int} + E_B,
\end{equation}
where $M$ is the mass of the atom and $\psi_m$ is the condensate wave
function for atoms in magnetic sublevel $m$, satisfying $\sum_m \int
d\bm{r} |\psi_m|^2 = N$ with $N$ being the total number of atoms.
We define the atomic number density
\begin{equation}
n = \sum_m |\psi_m|^2,
\end{equation}
and the spin density
\begin{equation} \label{F}
\bm{F} = \sum_{m m'} \psi_m^* \bm{f}_{m m'} \psi_{m'},
\end{equation}
where $\bm{f} = (f_x, f_y, f_z)$ is the vector of the spin-1 matrices.
The interaction energy $E_{\rm int}$ for the spin-1 atom has the
form~\cite{Ohmi,Ho}
\begin{equation}
E_{\rm int} = \frac{1}{2} \int d\bm{r} \left( c_0 n + c_1
\bm{F}^2 \right),
\end{equation}
where
\begin{subequations}
\begin{eqnarray}
c_0 & = & \frac{4 \pi \hbar^2}{M} \frac{a_0 + 2 a_2}{3}, \\
c_1 & = & \frac{4 \pi \hbar^2}{M} \frac{a_2 - a_0}{3},
\end{eqnarray}
\end{subequations}
with $a_S$ being the $s$-wave scattering length for two colliding atoms
with total spin $S$.
In the present paper, we consider the case of spin-1 $^{87}{\rm Rb}$
atoms, and take $a_0 = 101.8 a_{\rm B}$ and $a_2 = 100.4 a_{\rm
B}$~\cite{Kempen}, where $a_{\rm B}$ is the Bohr radius.
The linear and quadratic Zeeman energy $E_B$ under magnetic field $\bm{B}$
is given by
\begin{equation}
E_B = \int d\bm{r} \sum_{m m'} \psi_m^* \left[ -\frac{1}{2} \mu_{\rm B}
\bm{B} \cdot \bm{f} + \frac{\mu_{\rm B}^2}{4 E_{\rm hf}} (\bm{B} \cdot
\bm{f})^2 \right]_{m m'} \psi_{m'},
\end{equation}
where $\mu_{\rm B}$ is the Bohr magneton, $E_{\rm hf}$ is the hyperfine
splitting energy, and $1/2$ in the first term is the Land\'{e} g-factor.
The mean-field dynamics of the system is thus described by the
multicomponent Gross-Pitaevskii (GP) equation,
\begin{subequations}
\label{GP}
\begin{eqnarray}
i \hbar \frac{\partial \psi_{\pm 1}}{\partial t} & = & \left(
-\frac{\hbar^2}{2 M} \nabla^2 + V \mp \frac{1}{2} \mu_{\rm B} B + q + c_0
n \right) \psi_{\pm 1} \nonumber \\
& & + c_1 \left( \frac{1}{\sqrt{2}}
F_\mp \psi_0 \pm F_z \psi_{\pm 1} \right), \label{GP1} \\
i \hbar \frac{\partial \psi_0}{\partial t} & = & \left( -\frac{\hbar^2}{2
M} \nabla^2 + V + c_0 n \right) \psi_0 \nonumber \\
& & + \frac{c_1}{\sqrt{2}} \left(  F_+ \psi_1 + F_- \psi_{-1} \right),
\end{eqnarray}
\end{subequations}
where  $q = (\mu_{\rm B} B)^2 / (4 E_{\rm hf})$, $F_\pm = F_x \pm i F_y$,
and the magnetic field is assumed to be in the $z$ direction.

The linear Zeeman terms $\mp \mu_{\rm B} B / 2$ in Eq.~(\ref{GP1}) only
rotate the spin around the $z$ axis at the Larmor frequency.
Going onto the rotating frame of reference by setting $\psi_{\pm 1}
\rightarrow e^{\mp i \mu_{\rm B} B t / (2 \hbar)} \psi_{\pm 1}$ and $F_\pm
\rightarrow e^{\pm i \mu_{\rm B} B t / (2 \hbar)} F_\pm$, we find that the
linear Zeeman terms can be eliminated.

\subsection{Ground state and excitation spectrum}

We briefly review the ground state and excitation spectrum for the
homogeneous case~\cite{Stenger,Murata}.
For spin-1 $^{87}{\rm Rb}$ atoms, $c_1$ is negative, and the ground-state
phase diagram is given as Fig.~\ref{f:phase}, where $p = \mu_{\rm B} B / 2
+ p_0$ with $p_0$ being the Lagrange multiplier~\cite{Stenger}.
Here, $p_0$ is introduced to set $F_z$ at a prescribed value, since the
total magnetization in the $z$ direction is conserved.
When the quadratic Zeeman energy $q$ is larger than $2 |c_1| n$, the polar
phase, $(\psi_1, \psi_0, \psi_{-1}) = \sqrt{n} (0, 1, 0)$, is the ground
state.
Between the polar and ferromagnetic phases, there is a broken-axisymmetry
phase shown as the shaded region in Fig.~\ref{f:phase}.
The order parameter in this phase is given by~\cite{Murata}
\begin{subequations}
\label{baxi}
\begin{eqnarray}
\psi_{\pm 1} & = & (q \pm p) \sqrt{\frac{p^2 + 2 |c_1| n q - q^2}{8 |c_1|
n q^3}} \sqrt{n} e^{i \chi_{\pm 1}}, \\
\psi_0 & = & \sqrt{\frac{(q^2 - p^2)(p^2 + 2 |c_1| n q + q^2)}{4 |c_1| n
q^3}} \sqrt{n} e^{i (\chi_1 + \chi_{-1}) / 2}, \nonumber \\
\end{eqnarray}
\end{subequations}
where $\chi_{\pm 1}$ are arbitrary phases of the $m = \pm 1$ states.
We note that spin vector $\bm{F}$ in this phase tilts against the
direction of the magnetic field, breaking the axisymmetry
spontaneously~\cite{Murata}.
The phase boundaries are given by $q = |p|$ and $2 |c_1| n q - q^2 + p^2 =
0$, across which the system undergoes the second-order phase transition.

In Sec.~\ref{s:dynamics}, in an attempt to study the Berkeley experiment
we will examine the magnetization dynamics from the initial $m = 0$ state.
In order to understand the dynamics qualitatively, we consider the
Bogoliubov excitations from the $m = 0$ state for the homogeneous case.
Solving the Bogoliubov-de Genne equation, we obtain the excitation
spectrum as~\cite{SaitoA1,Murata}
\begin{eqnarray}
E_0 & = & \sqrt{\varepsilon_k (\varepsilon_k + 2 c_0 n)},
\label{E0} \\
E_{\pm 1} & = & \mp \frac{1}{2} \mu_{\rm B} B + \sqrt{(\varepsilon_k + q)
(\varepsilon_k + q + 2 c_1 n)}, \label{E1}
\end{eqnarray}
where $\varepsilon_k = \hbar^2 k^2 / (2 M)$ with $k$ being the wave
number.
The mode given in Eq.~(\ref{E0}) involves only the $m = 0$ state, and can
be regarded as a phonon mode, whose excitation energy is always real and
positive.
The two modes given in Eq.~(\ref{E1}) are magnon modes, which transfer the
atoms from the $m = 0$ state to the $m = \pm 1$ states.
The excitation energies $E_{\pm 1}$ are complex for $-q < \varepsilon_k <
-2 c_1 n - q$.
Therefore, when $c_1$ is negative and $q < 2 |c_1| n$, the $m = 0$ state
is dynamically unstable against long-wavelength excitations of the magnon
modes.
For $q < |c_1| n$, the most unstable wavelength is 
\begin{equation} \label{lambda}
\lambda_{\rm mu} = \frac{h}{\sqrt{2 M (|c_1| n - q)}},
\end{equation}
and the corresponding imaginary part of $E_{\pm 1}$ is $|c_1| n$, giving
a characteristic time scale for the dynamical instability as
\begin{equation} \label{tau}
\tau_{\rm mu} = \frac{\hbar}{|c_1| n}.
\end{equation}
For a larger magnetic field satisfying $|c_1| n \leq q < 2 |c_1| n$, the
most unstable wave number is $k = 0$, and therefore the system tends to
magnetize uniformly.
The corresponding time scale is given by $\hbar / [q |q + 2 c_1
n|]^{1/2}$.

The above Bogoliubov analysis is valid only when the system is homogeneous
and deviations from the initial state are small.
To investigate regions beyond such restrictions, we will solve full GP
equation (\ref{GP}) numerically in Sec.~\ref{s:dynamics}, and analyze the
detailed magnetization dynamics of the trapped system.

\subsection{Topological defects in spinor BECs}
\label{s:topological}

Quantized vortices and dark solitons in scalar BECs are topological
defects, in which the density vanishes due to the topological constraint
on the phase of the wave function.
We relate these topological defects in scalar BECs to spin vortices and
domain walls in spinor BECs, in which local magnetization vanishes.

We first note that throughout the spin-exchange dynamics the total density
$n = |\psi_{-1}|^2 + |\psi_0|^2 + |\psi_1|^2$ remains almost constant
because $c_0 \gg |c_1|$.
Since the transverse magnetization develops from the $m = 0$ state, we
also assume that $|\psi_{-1}|$ = $|\psi_1|$ and phases $\chi_m$ of
$\psi_m$ are related to each other by $\chi_{\pm 1} = \chi_0 \pm
\alpha(\bm{r}, t)$, where $\alpha$ is an arbitrary function.
Substituting these relations in the GP equation (\ref{GP}), we can
eliminate $\psi_{-1}$ and $\psi_0$, obtaining
\begin{equation} \label{singleGP}
i \hbar \frac{\partial \psi_1}{\partial t} = \left( -\frac{\hbar^2}{2 M}
\nabla^2 + q + c_0 n + 2 c_1 n - 4 c_1 |\psi_1|^2 \right) \psi_1,
\end{equation}
where we drop the trapping potential and the linear Zeeman term for
simplicity.
Equation~(\ref{singleGP}) is the single-component GP equation with a
repulsive interaction ($-4 c_1 > 0$).

\subsubsection{Polar-core vortices}
\label{s:pcv}

The general form of a spin-vortex state is given by
\begin{equation} \label{gensv}
\psi_m(\bm{r}) = e^{i \gamma_m} e^{i c_m \phi} f_m(r_\perp),
\end{equation}
where $\phi$ is the azimuthal angle, $\gamma_m$'s are global phases, 
$c_m$'s are integers, and $f_m$'s are real functions of $r_\perp = (x^2 +
y^2)^{1/2}$.
In Eq.~(\ref{gensv}), we assume that the system is uniform in the $z$
direction and that the vortex core is located on the $z$ axis.
For example, the Mermin-Ho texture has topological charges $c_1 = 0$, $c_0
= 1$, and $c_{-1} = 2$.

The solution to Eq.~(\ref{singleGP}) with a singly-quantized vortex is
given by
\begin{equation} \label{pcv}
\begin{array}{l}
\psi_1 = e^{\pm i \phi + i \gamma} f_1(r_\perp), \\
\psi_0 = f_0(r_\perp), \\
\psi_{-1} = e^{\mp i \phi - i \gamma} f_1(r_\perp),
\end{array}
\end{equation}
where $\gamma$ is an arbitrary phase and $f_0^2(r_\perp) + 2
f_1^2(r_\perp) = n$.
Since the $m = \pm 1$ states have topological defects, $f_1(r_\perp)$
vanishes at $r_\perp = 0$, and the core is then occupied by the $m = 0$
(polar) state.
The state given by Eq.~(\ref{pcv}) is therefore referred to as a
polar-core vortex.
Equation (\ref{singleGP}) indicates that $f_1$ is proportional to $r_\perp
/ \xi_s$ near $r_\perp = 0$, where
\begin{equation} \label{xis}
\xi_s = \frac{\hbar}{\sqrt{M (2 |c_1| n - q)}},
\end{equation}
and hence the size of the vortex core is characterized by $\xi_s$.

The spin density of Eq.~(\ref{pcv}) is given by
\begin{equation} \label{fp}
F_+ = \sqrt{2} \left( \psi_1^* \psi_0 + \psi_0^* \psi_{-1} \right)
= 2 \sqrt{2} e^{\mp i \phi - i \gamma} f_1(r_\perp) f_0(r_\perp),
\end{equation}
$F_z = 0$.
Equation~(\ref{fp}) shows that the spin vector circulates around the $z$
axis with phase winding $\mp 1$ and vanishes at $r_\perp = 0$.
In fact, the spin current
\begin{equation}
\bm{J}_{\rm spin}^{x,y,z} = \frac{\hbar}{2 M i} \sum_{m,m'} \left[
	\psi_m^* {F}_{m m'}^{x,y,z} \bm{\nabla} \psi_{m'} - (\bm{\nabla}
	\psi_m^*) F_{m m'}^{x,y,z} \psi_{m'} \right]
\end{equation}
for Eq.~(\ref{pcv}) is calculated to be $\bm{J}_{\rm spin}^x = \bm{J}_{\rm
spin}^y = 0$ and
\begin{equation}
\bm{J}_{\rm spin}^z = \frac{2 \hbar}{M r_\perp} f_1^2(r_\perp)
\bm{e}_\phi,
\end{equation}
where $\bm{e}_\phi$ is a unit vector in the azimuthal direction.
On the other hand, state (\ref{pcv}) has no mass current,
\begin{equation}
\bm{J}_{\rm mass} = \frac{\hbar}{2 M i} \sum_{m} \left[ \psi_m^*
	\bm{\nabla} \psi_m - (\bm{\nabla} \psi_m^*) \psi_m \right],
\end{equation}
and has no orbital angular momentum.

\subsubsection{Domain walls}
\label{s:darksoliton}

Equation~(\ref{singleGP}) has a dark-soliton solution as
\begin{equation} \label{dark}
\psi_1 = e^{-i \mu t / \hbar} \sqrt{\frac{2 |c_1| n - q}{8 |c_1|}}
\tanh \frac{z}{\sqrt{2} \xi_s},
\end{equation}
where $\mu = c_0 n - |c_1| n + q / 2$ and we assume that the planar dark
soliton is located at $z = 0$.
This solution describes a domain wall at $z = 0$, since the spin vector
$\bm{F}$ vanishes at $z = 0$ and asymptotically approaches a constant
vector with opposite directions for $z \rightarrow \pm \infty$.
The magnitude of transverse magnetization $|F_+|$ is proportional to
\begin{equation} \label{darkf}
\tanh \frac{z}{\sqrt{2} \xi_s} \sqrt{1 - \frac{1}{2} \tanh^2
\frac{z}{\sqrt{2} \xi_s}}.
\end{equation}
Thus, the width of the domain wall is of the same order as the size of the
spin vortex.

The planar dark soliton is known to be dynamically unstable against
``snake instability''~\cite{Kuz,Law}.
This instability arises for the wavelength longer than the critical
wavelength,
\begin{equation} \label{lamcr}
\lambda_{\rm cr} = \frac{h}{\sqrt{M |c_1| n}}.
\end{equation}
After the distortion by the snake instability, the dark soliton transforms
into vortex-antivortex pairs~\cite{Law}.
This phenomenon has been observed in a nonlinear optical
medium~\cite{Mamaev} and in a two-component BEC~\cite{Dutton}.
In the spinor BEC, an analogous instability causes formation of polar-core
vortex-antivortex pairs.
We will show in the next section that spin vortices are generated from
domain walls by this mechanism.

\section{Quenching dynamics of the trapped system}
\label{s:dynamics}

\subsection{The Berkeley experiment}
\label{s:experiment}

In this section, we will numerically simulate the magnetization dynamics
in a situation corresponding to the Berkeley experiment~\cite{Sadler} and
compare the results.
We first briefly review the experiment.

The trapping frequencies of the optical potential used in the experiment
are given by $(\omega_x, \omega_y, \omega_z) = 2 \pi \times (56, 350,
4.3)$ Hz, and the system is effectively quasi-two dimensional (2D) in the
$x$-$z$ plane.
The atoms prepared in the $m = -1$ state are transferred to the $m = 0$
state by rf field, and the magnetic field of $B = 2$ G is applied in the
$z$ direction.
The number of spin-1 $^{87}{\rm Rb}$ atoms in the BEC is $2.1 \times
10^6$ with a peak density of $n = 2.8 \times 10^{14}$ ${\rm cm}^{-3}$.
These conditions give $q / (|c_1| n) \simeq 28.4$, and therefore
the prepared state is stable (see Fig.~\ref{f:phase}).
The residual component in each of the $m = \pm 1$ states is less than
0.3\%.

The strength of the magnetic field is then suddenly decreased to 50 mG,
which corresponds to $q / (|c_1| n) \simeq 0.018$.
It follows from Fig.~\ref{f:phase} that the $m = 0$ state is no longer the
ground state and spontaneous magnetization emerges.
From $t = 50$ ms to 100 ms, the transverse ($x$-$y$ direction)
magnetization grows exponentially with the time constant of 15 ms.
The spin vector varies in space and points in various random directions,
and complicated spin textures can be observed.
The polar-core vortices are identified in about one-third of the snapshots
of the spin distribution, and sometimes several vortices coexist in a
single sample.
The correlation function for the transverse magnetization,
\begin{equation} \label{GT}
G_T(\delta\bm{r}) = {\rm Re} \left[ \frac{\int d\bm{r} F_-(\bm{r})
F_+(\bm{r} + \delta \bm{r})}{\int d\bm{r} n(\bm{r}) n(\bm{r} + \delta
\bm{r})} \right],
\end{equation}
oscillates in both $x$ and $z$ directions.
The wavelength of the oscillation in the $x$ direction is $\sim 10$ $\mu
{\rm m}$ and that in the $z$ direction is $\sim 50$ $\mu {\rm m}$.
The longitudinal magnetization $G_L(0)$ shows no significant change within
$t < 300$ ms, where
\begin{equation} \label{GL}
G_L(\delta\bm{r}) = \frac{\int d\bm{r} F_z(\bm{r})
F_z(\bm{r} + \delta \bm{r})}{\int d\bm{r} n(\bm{r}) n(\bm{r} + \delta
\bm{r})}.
\end{equation}

The quantities discussed in Sec.~\ref{s:meanfield} and observed in the
experiment are summarized in Tab.~\ref{tab}.
\begin{table}
\caption{Length and time scales obtained by the mean-field theory and
those observed in the Berkeley experiment~\cite{Sadler}.
The last row of the right column is blank, since it has not yet been
observed.
}
\label{tab}
\begin{ruledtabular}
\begin{tabular}{cc}
mean-field theory & experiment \\
\hline
$\lambda_{\rm mu}$ in Eq.~(\ref{lambda}) & 
wavelength of oscillation in $G_T(\delta\bm{r})$ \\
$\simeq 15.1$ $\mu{\rm m}$ &
$\sim 10$ $\mu{\rm m}$ (in $x$) \\
&
$\sim 50$ $\mu{\rm m}$ (in $z$) \\
\hline
$\tau_{\rm mu}$ in Eq.~(\ref{tau}) &
time constant of $G_T(0)$  \\
$\simeq 16$ ms &
$\simeq 15$ ms \\
\hline
$\xi_s$ in Eq.~(\ref{xis}) &
size of vortex core \\
$\simeq 2.4$ $\mu{\rm m}$ &
$\simeq 3$ $\mu{\rm m}$ \\
\hline
$\lambda_{\rm cr}$ in Eq.~(\ref{lamcr}) &
length scale of vortex-antivortex pair \\
$\simeq 21$ $\mu{\rm m}$ & 
?
\end{tabular}
\end{ruledtabular}
\end{table}

\subsection{Numerical method}
\label{s:method}

We perform full 3D numerical calculations of the GP equation (\ref{GP})
using the Crank-Nicolson scheme with a typical grid size of $\simeq 0.4$
$\mu {\rm m}$.
In the numerical calculations, we ignore the linear Zeeman terms in the GP
equation for the reason mentioned below Eq.~(\ref{GP}).
We first prepare the ground state in the $m = 0$ state, $\psi_0 =
\psi_{\rm ini}$, by the imaginary-time propagation of the GP equation
(\ref{GP}) with $\psi_{\pm 1} = 0$.

If the initial populations of the $m = \pm 1$ states are exactly zero, no
spin-exchange dynamics follow within the GP equation.
We therefore add small initial seeds to the $m = \pm 1$ states to trigger
the spin-exchange dynamics.
Possible physical origins of the initial seed include
(i) residual atoms due to imperfections in the rf transfer,
(ii) quantum fluctuations, and
(iii) thermal components.
In the experiment, although the residual fraction in each of the $m = \pm
1$ states is suppressed below 0.3\%~\cite{Sadler}, this upper limit
corresponds to 6300 atoms, which is still large enough to significantly
affect the subsequent dynamics.
The spatial distribution of the residual atoms should reflect that of the
condensate $\psi_0 = \psi_{\rm ini}$, while its phase and magnitude can
fluctuate spatially due to experimental noise.
Quantum fluctuations trigger the spin-exchange dynamics even if the
initial populations in the $m = \pm 1$ states are exactly zero, since
the $\hat\psi_1^\dagger \hat\psi_{-1}^\dagger \hat\psi_0^2$ term in the
second quantized Hamiltonian transfers the $m = 0$ population to the $m =
\pm 1$ states.
The quantum fluctuations can be taken into account by random noises in the
initial state~\cite{Norrie}.
The thermal component also triggers the growth of the $m = \pm 1$ states
having phase fluctuations.

In order to find an appropriate initial seed to reproduce the experimental
results and capture the essential mechanism that triggers the
magnetization, we will examine three kinds of initial seeds which reflect
the shape of the condensate and various types of noise.
We will see that the dynamics crucially depend on the nature of the
initial seed.

\subsection{Initial seed without noise}
\label{s:uniform}

In the Berkeley experiment the atoms are first prepared in the $m = -1$
state and then transferred to the $m = 0$ state.
We suppose that the transfer is imperfect and a small fraction is left in
the $m = -1$ state.
Thus we assume the initial state to be
\begin{equation} \label{uniform}
\begin{array}{l}
\psi_1 = 0, \\
\psi_0 = \sqrt{1 - \varepsilon^2} \psi_{\rm ini}, \\
\psi_{-1} = \varepsilon \psi_{\rm ini},
\end{array}
\end{equation}
where $\varepsilon$ is a small constant.
We take $\varepsilon = 0.05$ and hence the initial population in the $m =
-1$ state is 0.25\%, which is consistent with the experimental condition
($<$ 0.3\%).

\begin{figure}[tb]
\includegraphics[width=8.4cm]{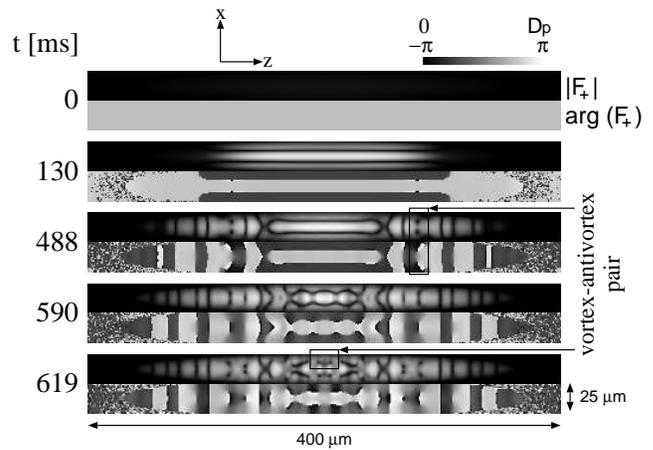}
\caption{
Magnitude and direction of the spin at $y = 0$ for the initial condition
in Eq.~(\ref{uniform}) with $\varepsilon = 0.05$.
The upper panels show transverse spin density $|F_+|$ and the lower
panels show its phase ${\rm arg}(F_+)$.
The gray scale represents the density from 0 to the peak value $D_{\rm p}
= n(\bm{r} = \bm{0})$ for the upper panels, and the phase from $-\pi$ to
$\pi$ for the lower panels.
The field of view of each panel is 400 $\times$ 25 $\mu{\rm m}$.
}
\label{f:uniform}
\end{figure}

Figure~\ref{f:uniform} shows snapshots of the transverse spin density
$|F_+|$ and its direction ${\rm arg}(F_+)$.
The spin-density profile at $t = 130$ ms exhibits three magnetic domains,
where the middle one magnetizes in the $x$ direction and the two side
ones in the $-x$ direction.
The width of each domain $\simeq 8$ $\mu{\rm m}$ roughly equals half of
the most unstable wavelength (\ref{lambda}).
As time proceeds, new domains are formed at both ends of the cigar-shaped
trap, whose alignment is perpendicular to the central domains (see the
snapshot at $t = 488$ ms in Fig.~\ref{f:uniform}).
Some of them become unstable and spin vortex-antivortex pairs are produced
(enclosed by the square at $t = 488$ ms in Fig.~\ref{f:uniform}).
Afterward, the two central domain walls become wavy ($t = 590$ ms), which
is followed by the formation of spin vortex-antivortex pairs (enclosed by
the square at $t = 619$ ms in Fig.~\ref{f:uniform}).
Around the spin vortex the direction of the spin vector rotates by $2\pi$
in the $x$-$y$ plane, and the core is occupied by the $m = 0$ component.
This topological defect is therefore the polar-core vortex discussed in
Sec.~\ref{s:pcv}.
Thus, the domain walls are first formed and the polar-core vortices then
develop from the domain walls.

We show that the dark-soliton picture discussed in
Sec.~\ref{s:darksoliton} describes the dynamics shown in
Fig.~\ref{f:uniform} very well.
The domain walls generated in Fig.~\ref{f:uniform} can be regarded as dark
solitons according to Eqs.~(\ref{singleGP}) and (\ref{dark}).
Fitting $|F_+|$ with Eq.~(\ref{darkf}), we find the width of the domain
walls along the $z$ axis at $t = 130$ ms is $\simeq 2$ $\mu{\rm m}$, and
$\simeq 3.3$ $\mu{\rm m}$ for those along the $x$ axis at $t > 420$ ms.
The latter is in good agreement with $\sqrt{2} \xi_s \simeq 3.4$ $\mu{\rm
m}$ and the former is somewhat smaller probably due to the influence of
the trapping potential.
The snake instability manifests itself as the wavy domain walls at $t =
590$ ms with a wavelength $\simeq 19$ $\mu{\rm m}$, which is roughly equal
to Eq.~(\ref{lamcr}).

\begin{figure}[tb]
\includegraphics[width=8.4cm]{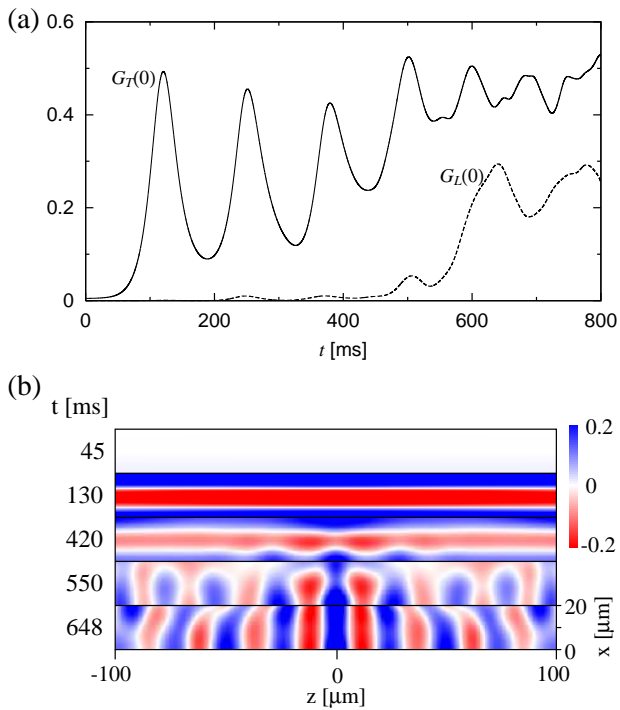}
\caption{
(Color) (a) Time evolution of $G_T(0)$ (solid curve) and $G_L(0)$ (dashed
curve) and (b) $G_T(x, z)$ for the initial condition given in
Eq.~(\ref{uniform}) with $\varepsilon = 0.05$.
}
\label{f:uniform2}
\end{figure}

Figure~\ref{f:uniform2} (a) shows time evolution of $G_T(0)$ and $G_L(0)$,
which indicate the degrees of magnetization in the $x$-$y$ and $z$
directions, respectively.
The transverse magnetization $G_T(0)$ exponentially grows in $\sim 100$ ms
with a time constant $\simeq 16$ ms, which is in good agreement with the
experimental observation of 15 ms.
The oscillation of $G_T(0)$ at the frequency of $\sim 8$ Hz is seen from
100 ms to 600 ms, which was not observed in the experiment.
The longitudinal magnetization $G_L(0)$ remains small for $t \lesssim 500$
ms.
The transverse correlation function $G_T(\delta\bm{r})$ is shown in
Fig.~\ref{f:uniform2} (b).
The stripe pattern in the $x$ direction emerges at $t \simeq 130$ ms,
which becomes unstable ($\simeq 420$ ms), transforming into the stripe in
the $z$ direction.

The initial growth of the magnetic domains staggered in the $x$ direction
is due to the anisotropy in the momentum distribution in the initial seed
(\ref{uniform}).
The Fourier component of the most unstable wavelength (\ref{lambda})
contained in the initial seed has the largest weight in the $x$
direction.
Since the domain walls staggered in the $x$ direction in
Fig.~\ref{f:uniform} (b) were not observed in the experiment, the initial
seed in Eq.~(\ref{uniform}) does not correspond to that in the
experiment.

\subsection{Initial seed with white noise}
\label{s:noise}

Next we examine the case of white noise as an initial seed in the $m = \pm
1$ states as
\begin{equation} \label{noise}
\begin{array}{l}
\psi_1 = {\cal N} (f_3 + i f_4) f_{\rm env}, \\
\psi_0 = {\cal N} \psi_{\rm ini}, \\
\psi_{-1} = {\cal N} (f_1 + i f_2) f_{\rm env},
\end{array}
\end{equation}
where ${\cal N}$ is a normalization constant, $f_i$'s are random
numbers obeying the normal distribution $e^{-f^2 / (2 \sigma^2)} /
(\sqrt{2 \pi} \sigma)$, and $f_{\rm env}$ is an envelope function.
The random number is chosen independently on each grid.
We take $\sigma = 2 \times 10^{-4}$, and the initial population in each of
the $m = \pm 1$ states is $\simeq 0.23$\%.
The envelope function $f_{\rm env}$ is taken to be $\psi_{\rm ini}$.

\begin{figure}[tb]
\includegraphics[width=8.4cm]{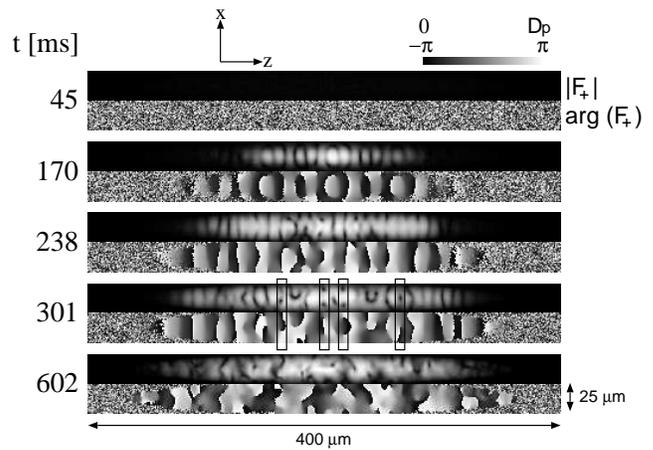}
\caption{
Magnitude and direction of the spin for the initial condition given in
Eq.~(\ref{noise}).
}
\label{f:noise}
\end{figure}

Figure~\ref{f:noise} shows the distributions of the magnitude and
direction of the spin for initial state (\ref{noise}).
In contrast to Fig.~\ref{f:uniform}, the domain structure staggered in the
$z$ direction first emerges at $t \simeq 170$ ms.
The size of the single domain is $\simeq 9$ $\mu{\rm m}$, which is roughly
the same as the domain width in Fig.~\ref{f:uniform} at $t = 130$ ms,
reflecting the fact that the domain size is set by the most unstable
length scale of the system.
Some of the domain walls then develop into the polar-core vortices as
shown in the squares in Fig.~\ref{f:noise}, and the system exhibits
complicated spin dynamics similar to the experimental results described in
Ref.~\cite{Sadler}.

\begin{figure}[tb]
\includegraphics[width=8.4cm]{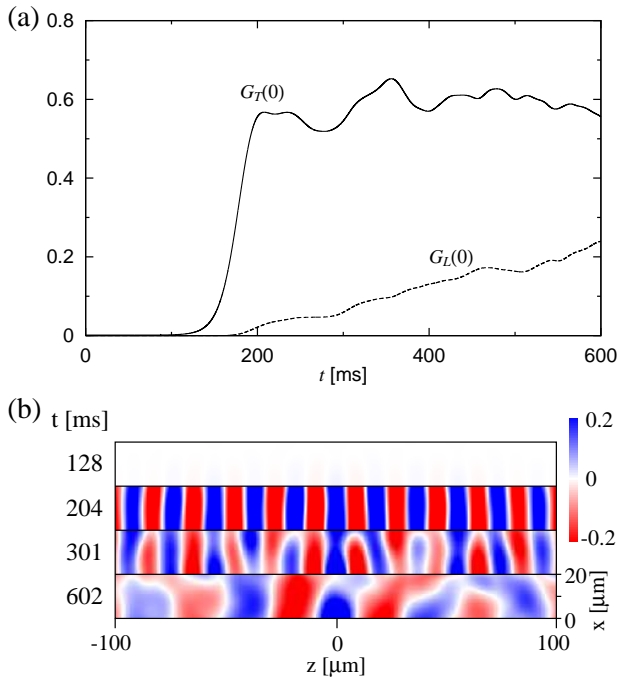}
\caption{
(Color) (a) Time evolution of $G_T(0)$ (solid curve) and $G_L(0)$ (dashed
curve) and (b) $G_T(x, z)$ for the initial condition given in
Eq.~(\ref{noise}).
}
\label{f:noise2}
\end{figure}

Time evolution of the transverse $G_T(0)$ and longitudinal $G_L(0)$ squared
magnetization is shown in Fig.~\ref{f:noise2} (a).
The time at which $G_T(0)$ rises in Fig.~\ref{f:noise2} (a) is later than
that in Fig.~\ref{f:uniform2}.
This is because the most unstable Fourier component in the initial seed is
smaller in the present case due to the broad momentum distribution of the 
white noise.
Figure~\ref{f:noise2} (b) shows the transverse correlation function
$G_T(\delta \bm{r})$.
The clear stripe pattern at $t = 204$ ms indicates the establishment of
the long-range correlation over $> 200$ $\mu{\rm m}$.
The stripe pattern is then distorted and its width becomes broader for $t
\gtrsim 600$ ms.

We note that the qualitative behaviors in Figs.~\ref{f:noise} and
\ref{f:noise2} are relatively insensitive to the envelope function $f_{\rm
env}$ of the initial seed in Eq.~(\ref{noise}).
We have used $f_{\rm env} = \psi_{\rm ini}$ based on the assumption that
the noise reflects the shape of the condensate.
However, we find that the results remain qualitatively the same even when
the envelope function is not multiplied.
We have also confirmed that the results shown above are insensitive to
grid size, despite the fact that a random number is initially assigned to
each grid and the momentum distribution of the initial noise depends on
the grid size.

\subsection{Initial seed with colored noise}
\label{s:cutoff}

The initial states (\ref{uniform}) and (\ref{noise}) examined in
Secs.~\ref{s:uniform} and \ref{s:noise} have been inadequate to reproduce
the experimental results.
In the experiments, the transverse correlation function $G_T(\delta
\bm{r})$ oscillates both in the $x$ and $z$ directions and the wavelength
of the oscillation is larger in the $z$ direction than that in the $x$
direction.
For the initial seed proportional to $\psi_{\rm ini}$ given in
Eq.~(\ref{uniform}), the domains staggered in the short axis first grow,
and for the white noise in Eq.~(\ref{noise}), the domains are staggered in
the long axis.
Therefore, we expect that the experimental result can be reproduced using
the initial seed that combines Eqs.~(\ref{uniform}) and (\ref{noise}),
i.e., anisotropy and randomness.
Since the white noise in Eq.~(\ref{noise}) induces the growth of the
domains along the $x$ direction (Fig.~\ref{f:noise}, $t = 170$ ms) and the
wavelength of the oscillation of $G_T(\delta \bm{r})$, $\simeq 18$ $\mu
{\rm m}$ (Fig.~\ref{f:noise2}, $t = 204$ ms), is shorter than that in the
experiment, $\simeq 50$ $\mu {\rm m}$, we cut off the short-wavelength
components from the white noise.
The initial state is thus given by
\begin{equation} \label{cutoff}
\begin{array}{l}
\psi_1 = 0, \\
\psi_0 = {\cal N} \psi_{\rm ini}, \\
\psi_{-1} = {\cal N} f_{\rm cutoff} f_{\rm env},
\end{array}
\end{equation}
where we produce the noise function $f_{\rm cutoff}$ from the white noise
by eliminating the Fourier components whose wavelengths are shorter than
$\lambda_{\rm cutoff}$.
The envelope function $f_{\rm env}$ is taken to be $\psi_{\rm ini}$.

\begin{figure}[tb]
\includegraphics[width=8.4cm]{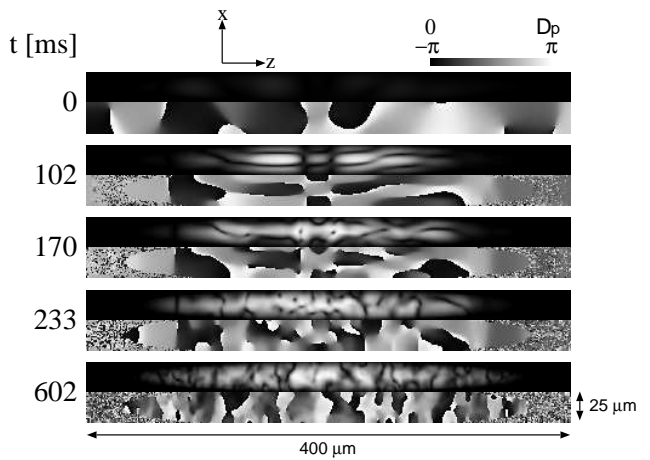}
\caption{
Magnitude and direction of the spin for the initial condition given in
Eq.~(\ref{cutoff}) with $\lambda_{\rm cutoff} = 60$ $\mu {\rm m}$.
}
\label{f:cutoff}
\end{figure}

\begin{figure}[tb]
\includegraphics[width=8.4cm]{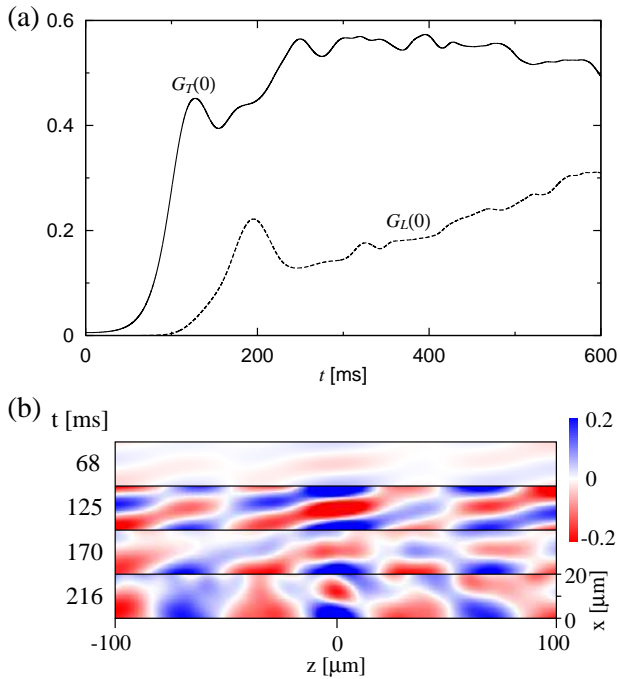}
\caption{
(Color) (a) Time evolution of $G_T(0)$ (solid curve) and $G_L(0)$ (dashed
curve) and (b) $G_T(x, z)$ for the initial condition given in
Eq.~(\ref{cutoff}) with $\lambda_{\rm cutoff} = 60$ $\mu {\rm m}$.
}
\label{f:cutoff2}
\end{figure}

The results for $\lambda_{\rm cutoff} = 60$ $\mu {\rm m}$ are shown in
Figs.~\ref{f:cutoff} and \ref{f:cutoff2}.
The behavior of $G_T(\delta \bm{r})$ shown in Fig.~\ref{f:cutoff2} (b) is
similar to that observed in the experiment~\cite{Sadler} in that the
wavelength of the oscillation in the $z$ direction is larger than that in
the $x$ direction.
This anisotropy originates from the envelope function $f_{\rm env} =
\psi_{\rm ini}$ of the initial seed in Eq.~(\ref{cutoff}).
The momentum distribution in the $x$ direction of $f_{\rm env}$ is broad
and the most unstable wavelength~(\ref{lambda}) grows in this direction,
while in the $z$ direction the typical wavelength in $G_T(\delta \bm{r})$
is determined by the cutoff wavelength $\lambda_{\rm cutoff}$ of the
noise.

We note that the qualitative behaviors shown in Fig.~\ref{f:cutoff2} (b)
are independent of the details of the envelope function $f_{\rm env}$ in
the initial seed.
The similar behaviors can be obtained as long as the size of the envelope
function is much larger than the unstable wavelength $\lambda_{\rm mu}$ in
the $z$ direction and is comparable to $\lambda_{\rm mu}$ in the $x$
direction.
In fact, when we use $f_{\rm env} = \psi_{\rm ini}^2$, qualitatively
similar results are obtained.

The behavior of the local transverse magnetization $G_T(0)$ in
Fig.~\ref{f:cutoff2} (a) is also in good agreement with the experimental
result.
It exponentially increases from $t = 50$ ms to $100$ ms with a time
constant $\simeq 15$ ms.
After $t = 100$ ms, $G_T(0)$ gradually increases from 0.4 to 0.6 until $t
= 300$ ms, which also captures the basic characteristics of the experimental
result.

As in Figs.~\ref{f:uniform} and \ref{f:noise}, Fig.~\ref{f:cutoff} shows
that the magnetic domains are first formed ($t = 102$ ms) followed by the
development of some of the domain walls into the polar-core vortices.
Thus, this process of spin-vortex formation appears rather universal.
We also found that the polar-core vortices drift in and out of the
condensate in the dynamics.

\subsection{Dependence of the dynamics on quench time and magnetic field}
\label{s:quench}

We have so far considered the case of sudden quench, i.e., the magnetic
field being suddenly reduced to 50 mG at $t = 0$.
If the time scale of the quench is longer than $\sim 100$ ms (time scale
in which $G_T(0)$ rises in Fig.~\ref{f:cutoff2}), the excitations are
expected to be suppressed because of the adiabatic theorem.

\begin{figure}[tb]
\includegraphics[width=8.4cm]{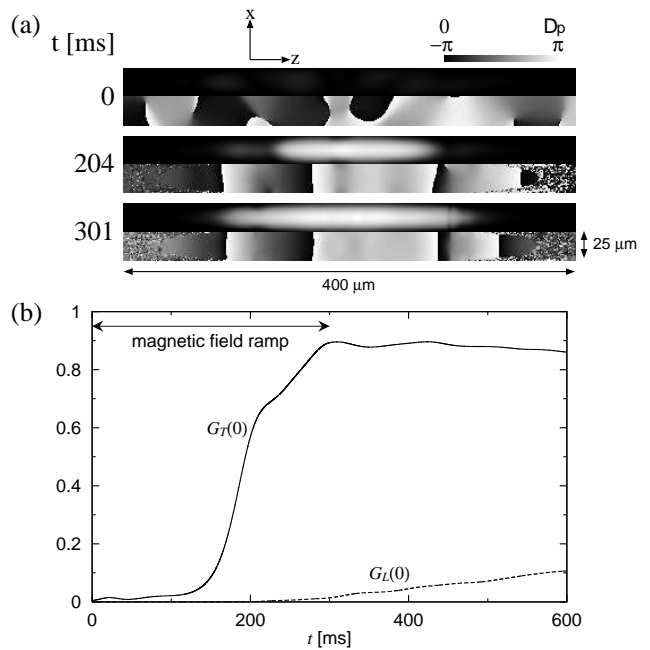}
\caption{
Time evolution of the system, in which the magnetic field is decreased
from 530 mG to 50 mG during the first 300 ms so that $q$ linearly
decreases.
The initial condition is the same as in Figs.~\ref{f:cutoff} and
\ref{f:cutoff2}.
}
\label{f:slowquench}
\end{figure}

Figure~\ref{f:slowquench} shows the time evolution for the slow quench,
where the initial magnetic field of 530 mG is reduced to 50 mG during 300
ms so that $q$ linearly decreases.
We find from Fig.~\ref{f:slowquench} (a) that the spin state has nearly a
single-domain structure and no spin vortices are created.
The transverse $G_T(0)$ and longitudinal $G_L(0)$ components of the
squared magnetization are shown in Fig.~\ref{f:slowquench} (b).
A large value of $G_T(0) \simeq 0.9$ is due to the absence of the spatial
spin structure.

\begin{figure}[tb]
\includegraphics[width=8.4cm]{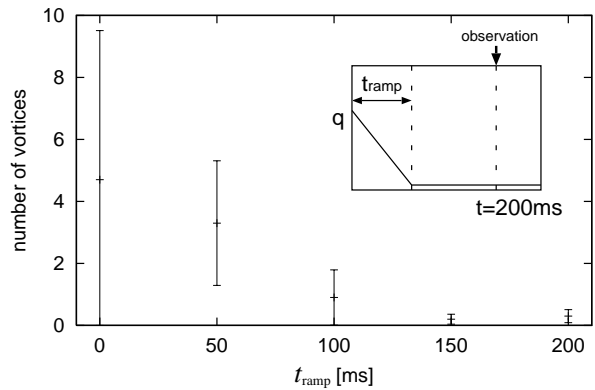}
\caption{
Ramp time dependence of the number of spin vortices at $t = 200$ ms.
The initial state is given in Eq.~(\ref{cutoff}) with $\lambda_{\rm
cutoff} = 60$ $\mu {\rm m}$ and the magnetic field is decreased from 530
mG to 50 mG so that $q$ is linearly ramped down during $t_{\rm ramp}$.
The plots and error bars represent the average and standard deviation with
respect to ten runs of simulations for different random numbers to
generate the noise.
}
\label{f:vortexnum}
\end{figure}

Figure~\ref{f:vortexnum} shows the dependence of the number of spin
vortices at $t = 200$ ms on ramp time.
The number of spin vortices decreases with an increase in the ramp time,
and the time scale beyond which the spin vortices no longer emerge is
given by $\sim 100$ ms.
We note that this time scale coincides with $h / (|c_1| n)$, indicating
that the energy scale for the creation of a spin vortex is given by $\sim
|c_1| n$.
As in the experiment, the number of vortices fluctuates from run to run.

The result seen in Fig.~\ref{f:vortexnum} reminds us of the Kibble-Zurek
mechanism~\cite{Kibble,Zurek} of the vortex creation in the quenched
system, which also depends on the quench time.
However, the relationship between the present phenomenon and the
Kibble-Zurek mechanism is not straightforward.
In the Kibble-Zurek mechanism, each domain is assumed to be created with
an independent phase.
In the present system, however, the magnetization at each position cannot
be independent of that of other positions, since there is the restriction
on the change in the total spin, which is clearly seen in the long-range
correlation in $G_T(\delta \bm{r})$.
Gaining a deeper understanding of the relationship between the spin-vortex
formation and the Kibble-Zurek mechanism constitutes an interesting and
challenging problem which merits further investigation.

\begin{figure}[tb]
\includegraphics[width=8.4cm]{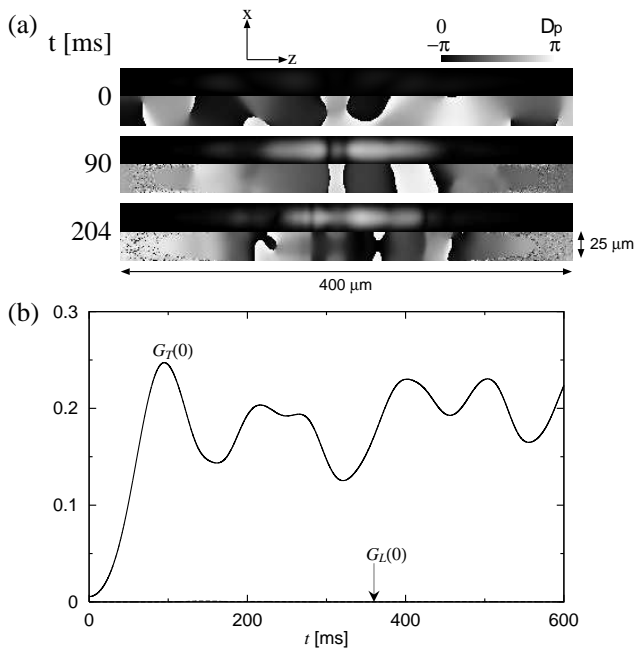}
\caption{
Time evolution of the system, in which the magnetic field is suddenly
decreased to 400 mG at $t = 0$.
The initial condition is the same as in Figs.~\ref{f:cutoff} and
\ref{f:cutoff2}.
}
\label{f:largem}
\end{figure}

The spin-vortex formation is also suppressed if the quench of the magnetic
field is made just below the critical strength of the magnetic field,
which is given by $q = 2 |c_1| n$ (see Fig.~\ref{f:phase}) and corresponds
to $B = 530$ mG for the peak density.
Figure~\ref{f:largem} shows the spin dynamics for $B = 400$ mG.
From Eq.~(\ref{E1}), the most unstable wavelength for this magnetic field
is $\lambda = \infty$, and the system is more unstable against excitations
with larger wavelengths.
In Fig.~\ref{f:largem} (a), therefore, only the long-wavelength modes are
excited and no spin vortex is formed.
In Fig.~\ref{f:largem} (b), the transverse squared magnetization $G_T(0)$
saturates around 0.2, since $|F_+|$ of the ground state is small near the
phase boundary between the broken-axisymmetry and polar phases.

\subsection{Total transverse magnetization}
\label{s:transverse}

We define the total magnetization as
\begin{equation}
\langle \bm{F} \rangle = \frac{1}{N} \int d\bm{r} \bm{F},
\end{equation}
where $\bm{F}$ is the spin density defined in Eq.~(\ref{F}).
If the quadratic Zeeman effect is absent, the $z$ component $\langle F_z
\rangle$ and the transverse component $|\langle F_+ \rangle|$ are 
conserved, and the vector $(\langle F_x \rangle, \langle F_y \rangle)$
rotates in the $x$-$y$ plane at the Larmor frequency determined by the
linear Zeeman energy.
In the presence of the quadratic Zeeman effect, not only $\langle \bm{F}
\rangle$ rotates in the $x$-$y$ plane but also $|\langle F_+ \rangle|$
changes with time.

\begin{figure}[tb]
\includegraphics[width=8.4cm]{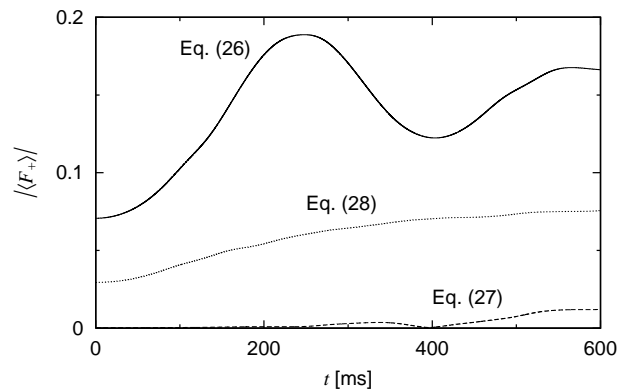}
\caption{
Time evolution of the transverse magnetization $| \langle F_+ \rangle | =
|\int d\bm{r} F_+ |$ for the dynamics in Fig.~\ref{f:uniform}
(solid curve), Fig.~\ref{f:noise} (dashed curve), and Fig.~\ref{f:cutoff}
(dotted curve).
}
\label{f:transverse}
\end{figure}

Figure~\ref{f:transverse} shows time evolution of $|\langle F_+ \rangle|$
for the initial conditions given by Eqs.~(\ref{uniform}), (\ref{noise}),
and (\ref{cutoff}), where the system is quenched by a decrease in the
magnetic field to 50 mG at $t = 0$ as in
Figs.~\ref{f:uniform}-\ref{f:cutoff2}.
For the initial condition (\ref{uniform}), $|\langle F_+ \rangle|$ has a
large initial value $\simeq 0.07$, and reaches a maximum value $\sim
0.19$.
For the initial condition (\ref{cutoff}), $|\langle F_+ \rangle|$
monotonically increases to twice the initial value during 600 ms.
Thus the quadratic Zeeman effect generates the transverse component in the
total magnetization.

The transverse magnetization $|\langle F_+ \rangle|$ may also be changed
by the dipole-dipole interaction, since it couples the spin angular
momentum with the orbital angular momentum~\cite{Kawaguchi,Santos}.
The effect of the dipole-dipole interaction on the transverse
magnetization merits further study.

\section{Conclusions}
\label{s:conclusion}

We have studied the spontaneous magnetization and spin-texture formation
of a spin-1 $^{87}{\rm Rb}$ BEC, where the initial state is the $m = 0$
stationary state $\psi_{\rm ini}$ plus a small seed in the $m = \pm 1$
states.

We have reproduced the polar-core spin-vortex formation observed in
the experiment, as shown in Figs.~\ref{f:uniform}, \ref{f:noise}, and
\ref{f:cutoff}.
Typically, the spin vortex is formed in two steps.
The magnetic domains are first formed, and then the domain walls transform
into the spin vortex-antivortex pairs.
This process of vortex formation appears to be universal regardless of
various initial conditions.

We have examined three kinds of initial seeds: the one proportional to
$\psi_{\rm ini}$, white noise, and colored noise.
The first two seeds produce the domain structures staggered in the short
and long axes, respectively (Figs.~\ref{f:uniform}-\ref{f:noise2}).
The magnetization developed from the third seed has both characteristics
of the first two seeds, and the correlation function oscillates in both
the long and the short axes (Fig.~\ref{f:cutoff2} (a)), in qualitative
agreement with the Berkeley experiment~\cite{Sadler}.
This is due to the fact that the third seed has a broad momentum
distribution in the short axis, originating from the shape of $\psi_{\rm
ini}$, and long-wavelength fluctuations in the long axis.
From these results, we can conclude that the anisotropy and colored noise
in the initial seed are important to account for the experiment.
The time evolution of the transverse magnetization $G_T(0)$ is also in
close agreement with the experimental result (Fig.~\ref{f:cutoff2} (b)).

The number of spin vortices created in the magnetization depends on how
fast the magnetic field is quenched.
When the magnetic field is decreased slowly, the number of nucleated spin
vortices decreases (Figs.~\ref{f:slowquench} and \ref{f:vortexnum}).
We have also shown that the number of spin vortices decreases for the
magnetic field close to the critical value (Fig.~\ref{f:largem}).

The transverse component of the total magnetization $|\langle F_+
\rangle|$ is changed by the quadratic Zeeman effect, and can exceed twice
the initial value (Fig.~\ref{f:transverse}).

We have pointed out a close analogy between the topological defects in the
present system and those in a scalar BEC, and that the creation of spin
vortex-antivortex pairs from the domain walls is related to the
instability in the planar dark solitons.
It is of interest to investigate if the counterparts of vortex lattices,
multiply-quantized vortices, and gray solitons are generated in a
ferromagnetic spinor BEC.

\begin{acknowledgments}
This work was supported by Grants-in-Aid for Scientific Research (Grant
Nos.\ 17740263 and 17071005) and by the 21st Century COE programs on
``Coherent Optical Science'' and ``Nanometer-Scale Quantum Physics'' from
the Ministry of Education, Culture, Sports, Science and Technology of
Japan.
YK acknowledges support by the Japan Society for Promotion of Science
(Project No.\ 185451).
MU acknowledges support by a CREST program of the JST.
\end{acknowledgments}

\end{document}